\documentclass[aps,twocolumn,nofootinbib,preprintnumbers,superscriptaddress]{revtex4-1}
\usepackage{amsmath}
\usepackage{amssymb}
\usepackage{mathrsfs}
\usepackage{graphicx}
\usepackage{gensymb}
\usepackage{float}
\usepackage{bm} 
\usepackage{svg}
\usepackage[T1]{fontenc}
\usepackage{natbib}
\usepackage{braket}
\usepackage{gensymb}
\usepackage{epstopdf}
\usepackage{longtable}
\usepackage{multirow}
\usepackage{makecell}
\usepackage{siunitx}
\usepackage[normalem]{ulem}
\usepackage[colorlinks=true,citecolor=blue,linkcolor=blue,urlcolor=blue]{hyperref}
\makeatother
\usepackage{xcolor}

\begin{document}
\title{Enhanced Non-linear Response by Manipulating the Dirac Point in the (111) LaTiO$_3$/SrTiO$_3$ Interface}
\author{G. Tuvia}
\affiliation{School of Physics and Astronomy, Tel-Aviv University, Tel Aviv, 6997801, Israel}
\author{A. Burshtein}
\affiliation{School of Physics and Astronomy, Tel-Aviv University, Tel Aviv, 6997801, Israel}
\author{I. Silber}
\affiliation{School of Physics and Astronomy, Tel-Aviv University, Tel Aviv, 6997801, Israel}
\author{A. Aharony}
\affiliation{School of Physics and Astronomy, Tel-Aviv University, Tel Aviv, 6997801, Israel}
\author{O. Entin-Wohlman}
\affiliation{School of Physics and Astronomy, Tel-Aviv University, Tel Aviv, 6997801, Israel}
\author{M. Goldstein}
\affiliation{School of Physics and Astronomy, Tel-Aviv University, Tel Aviv, 6997801, Israel}
\author{Y. Dagan}
\email{Corresponding author: yodagan@tauex.tau.ac.il}
\affiliation{School of Physics and Astronomy, Tel-Aviv University, Tel Aviv, 6997801, Israel}
\begin{abstract}
Tunable spin-orbit interaction (SOI) is an important feature for future spin-based devices. In the presence of a magnetic field, SOI induces an asymmetry in the energy bands, which can produce non-linear transport effects ($V\sim I^2$). Here, we focus on such effects to study the role of SOI in the (111) LaTiO$_3$/SrTiO$_3$ interface. This system is a convenient platform for understanding the role of SOI since it exhibits a single-band Hall-response through the entire gate-voltage range studied. We report a pronounced rise in the non-linear longitudinal resistance at a critical in-plane field $H_{cr}$. This rise disappears when a small out-of-plane field component is present. We explain these results by considering the location of the Dirac point formed at the crossing of the spin-split energy bands. An in-plane magnetic field pushes this point outside of the Fermi contour, and consequently changes the symmetry of the Fermi contours and intensifies the non-linear transport. An out-of-plane magnetic field opens a gap at the Dirac point, thereby significantly diminishing the non-linear effects. We propose that magnetoresistance effects previously reported in interfaces with SOI could be comprehended within our suggested scenario. 
\end{abstract}
\maketitle

Rashba-type spin-orbit interaction (SOI) \cite{rashba1960properties} arises in systems lacking inversion symmetry, such as polar semiconductors \cite{ishizaka2011giant,ideue2017bulk} and interfaces or surfaces \cite{santander2014giant,lashell1996spin}. This interaction lifts the spin degeneracy, leading to the splitting of the electronic bands depending on their spin. Such spin-split systems exhibit a multitude of phenomena that hold great promise for spin-based devices \cite{manchon2015new}. Two-dimensional electron systems are especially interesting in this respect, as the magnitude of their SOI can be tuned by gating \cite{premasiri2019tuning}. For example, in SrTiO$_3$ based interfaces, SOI plays a role in phenomena such as weak anti-localization \cite{rakhmilevitch2010phase}, magnetoresistance effects \cite{shalom2010tuning}, superconductivity \cite{rout2017link}, and formation of Berry curvature dipoles \cite{lesne2023designing}, all of which can be tuned by gate. The (111) surface of SrTiO$_3$ is particularly intriguing due to its triangular structure, which further reduces the symmetry, leading to rich spin textures \cite{he2018observation,lesne2023designing}, and six-fold anisotropic magnetoresistance \cite{rout2017six}. 
\par
Non-linear transport effects ($V\sim I^2$) are useful probes for exploring quantum materials with broken inversion symmetry. The various components of the non-linear tensor contain rich information on a variety of phenomena such as skew scattering \cite{isobe2020high,he2021quantum}, superconducting fluctuations
\cite{itahashi2020nonreciprocal, itahashi2022giant}, and Berry curvature dipoles \cite{lesne2023designing}. Such phenomena may result in pronounced effects in the non-linear transport while being hard to detect in conventional linear measurements.
\par
For systems with SOI, a magnetic field causes the spin-split bands to move in momentum space perpendicular to the magnetic field and the direction in which inversion is broken. This asymmetric band structure can lead to a longitudinal non-linear resistance which is antisymmetric with respect to the magnetic field. This antisymmetric component can lend useful information regarding spin-texture and the band structure of the system. \cite{ideue2017bulk,he2018observation,he2019nonlinear,choe2019gate,li2021nonreciprocal}.

\begin{figure}[b]
\centering
\includegraphics[width=\linewidth]{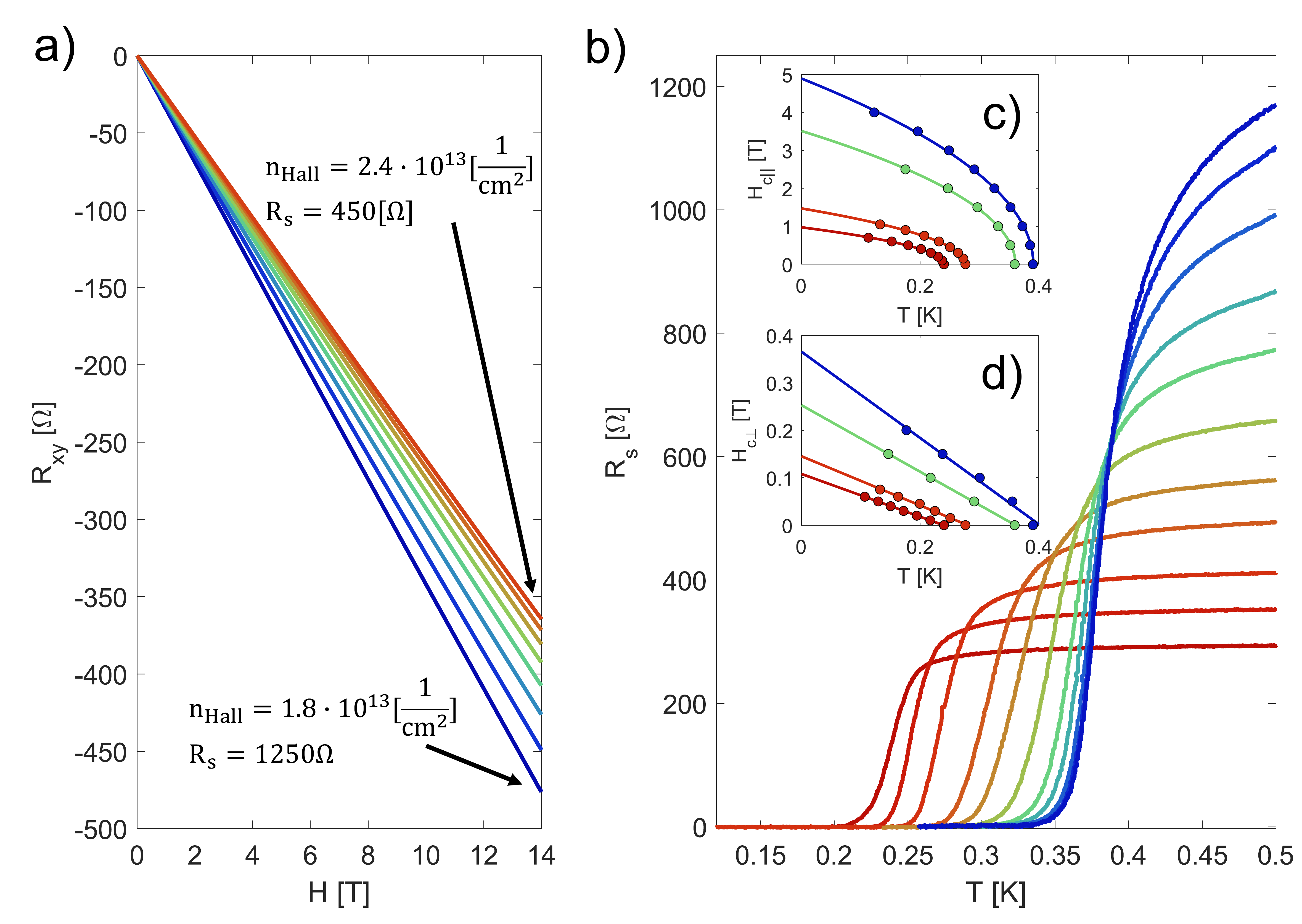}
\caption{Hall and superconductivity measurements. (a) Hall effect for various gate-voltages at 10 K, displaying a perfectly linear one-band behavior. (b) Superconducting transitions for various gates-voltages. (c)-(d) In-plane and out-of-plane superconducting critical fields at various temperatures and gates exhibiting a two-dimensional behavior.} 
\label{fig:SCandHall}
\end{figure}

At the interface between the Mott insulator LaTiO$_3$ and the band insulator SrTiO$_3$, the Ti$^{+3}$ and Ti$^{+4}$  valences hybridize \cite{ohtomo2002artificial,wang2020anomalous}. For the (001) interface, this region of hybridized Ti valences was shown to result in two-dimensional conductivity \cite{ohtomo2002artificial} and superconductivity \cite{biscaras2010two}. 

Here we study the electronic properties of the (111) LaTiO$_3$/SrTiO$_3$ interface. We focus on the longitudinal antisymmetric term of the non-linear resistivity tensor. We identify a field scale, $H_{cr}$, beyond which this non-linear component is enhanced. We present a simple model that associates $H_{cr}$ with the field at which a Rashba-Dirac point is excluded from the Fermi contour.

A thin layer of LaTiO$_3$ is deposited on a (111) SrTiO$_3$ substrate by pulsed laser deposition. Devices with current driven along various crystal axes are defined using photolithography (see supplemental material for details\cite{sm}). We begin by showing that for all gate voltages studied, the Hall resistance exhibits a linear, single-band-type dependence on the magnetic field with no significant deviations up to 14 T (Fig. \ref{fig:SCandHall}(a)). 

At low temperatures, the interface becomes superconducting (Fig. \ref{fig:SCandHall}(b)). We find that as the gate-voltage \textit{increases} (higher carrier density and lower sheet resistance), the superconducting transition temperature $T_c$ \textit{decreases}, typical of an ``overdoped'' regime. Throughout this work, results are plotted as a function of the sheet resistance instead of gate-voltage to account for the dependence of the resistance on the training history of the gate-voltage (see correspondence between the sheet resistance and the carrier density in Fig. S6 \cite{sm}). 

To demonstrate the two-dimensional nature of superconductivity, we present the in-plane and out-of-plane superconducting critical magnetic fields in Fig. \ref{fig:SCandHall}(c)-(d), respectively. For all gate-voltages studied, we observe a linear (square root) relation between the out-of-plane (in-plane) superconducting critical field and the temperature as expected for two-dimensional superconductivity. Furthermore, the superconducting coherence length ($\xi=30-55$ nm, depending on gate) and the upper bound for the thickness of the superconducting layer ($d=8-21$ nm) satisfy $\xi>d$ for all gate-voltages.

To measure the non-linear transport, we apply an alternating current and measure the longitudinal second harmonic voltage drop \cite{ideue2017bulk,he2018observation,choe2019gate}. We focus on antisymmetric terms $V_{2\omega}(H)=-V_{2\omega}(-H)$, where $H$ is the magnetic field. It is useful to define $\gamma=\frac{2R_{2\omega}}{R_0HI}$, where $R_{2\omega}=V_{2\omega}/I$ is the second-harmonic resistance, $R_0$ is the Ohmic resistance, $I$ is the current, $\gamma$ is the coefficient of the non-linear term in the resistance:

\begin{equation}\label{eq:gamma}
R(I,H)=R_0(1+\beta H^2+\gamma IH),
\end{equation}
and $\beta$ is the orbital magnetoresistance coefficient.  

\begin{figure}[htp]
\centering
\includegraphics[width=\linewidth]{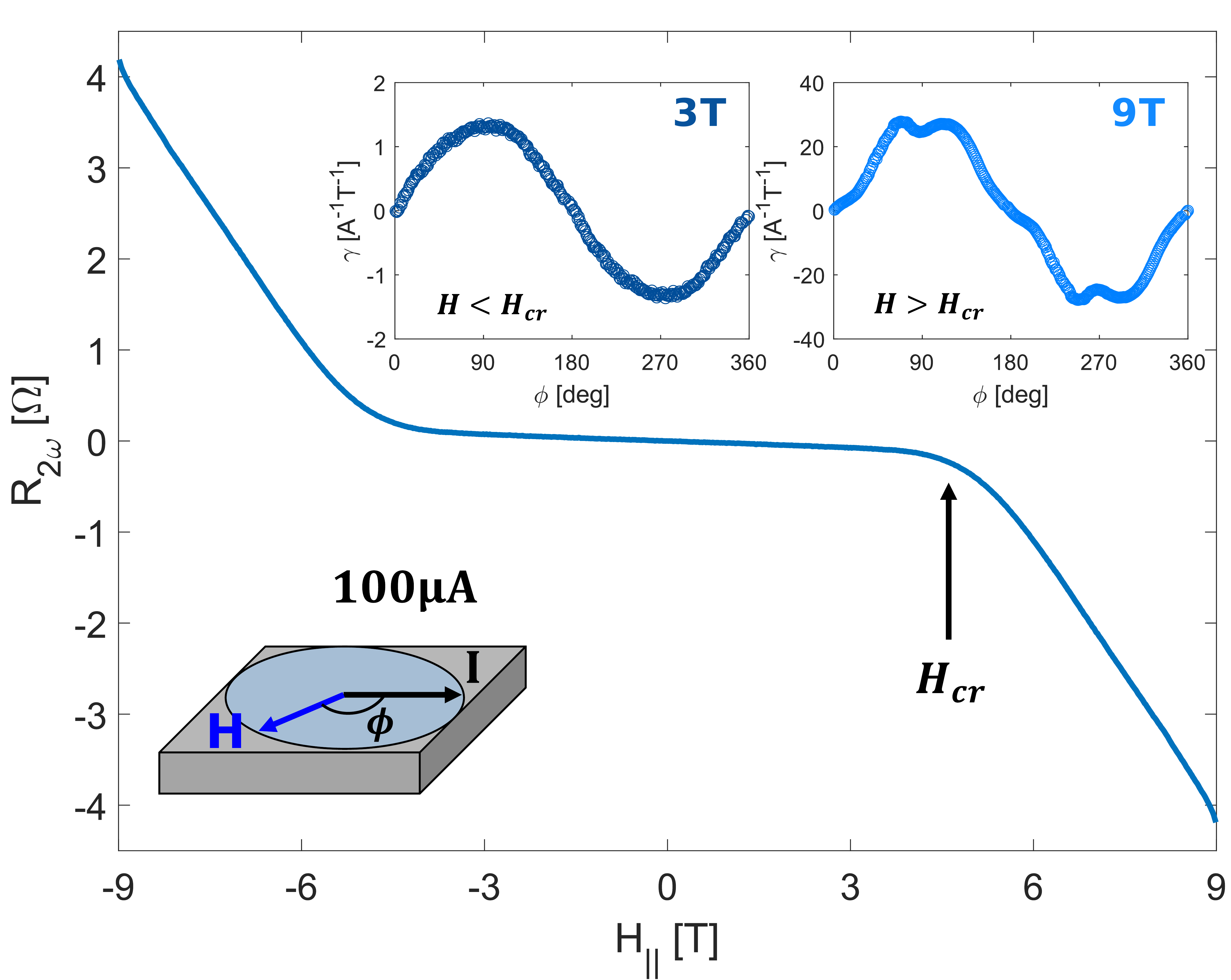}
\caption{The second-harmonic resistance versus the in-plane magnetic field at T=3 K and $R_s$=380 $\Omega$; $H_{cr}$ is indicated. Top inset: Dependence of $\gamma=\frac{2R_{2\omega}}{R_0HI}$ on the in-plane angle $\phi$. In the low-field regime, the effect depends only on the component of the field perpendicular to the current. In the high-field regime, the effect also depends on the in-plane field angle $\phi$. Bottom inset: Sketch of the sample, current, and magnetic field. In all such sketches throughout this work, the SrTiO$_3$ is facing up, while the LaTiO$_3$ is facing down.} 
\label{fig:InP}
\end{figure}

The second-harmonic resistance of the interface exhibits a pronounced enhancement beyond a critical (in-plane) transverse magnetic field $H_{cr}$ (Fig. \ref{fig:InP}). For the low-field regime ($H<H_{cr}$), the second-harmonic effect is small and depends linearly on the transverse magnetic field. This behavior is further illustrated in the low-field inset, where $\gamma$ has a sinusoidal dependence on the in-plane angle $\phi$, defined as the angle between the in-plane magnetic field and the current (bottom inset). However, as the magnetic field exceeds the critical magnetic field ($H>H_{cr}$), $\gamma$ increases substantially and exhibits oscillations depending on $\phi$ (see high-field inset). As can be seen in the supplemental material Fig. S2 \cite{sm}, these high-field oscillations depend on the magnitude of the magnetic field and on direction of the current with respect to the crystal axes.

\begin{figure*}
\includegraphics[width=\textwidth]{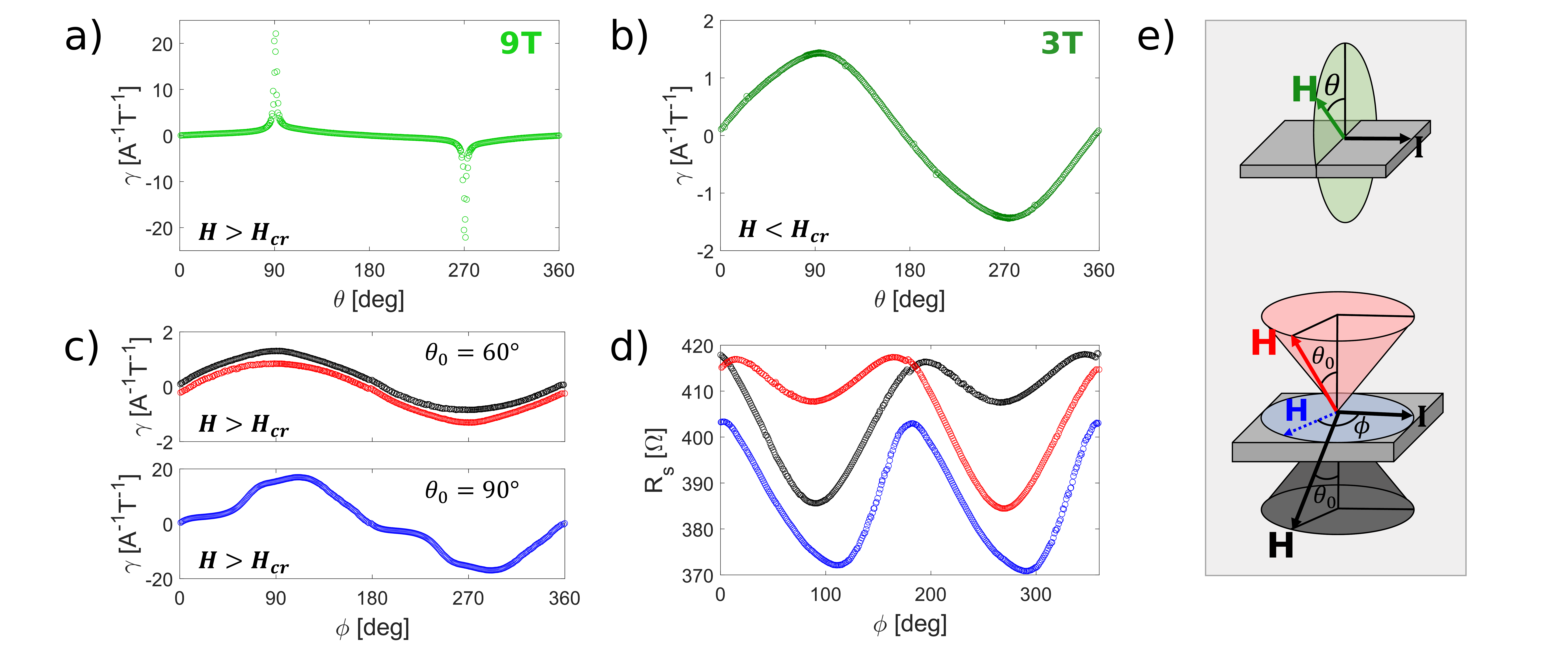}
  \caption{Effects of the out-of-plane magnetic field. (a,b) $\gamma$ is shown as a function of the out-of-plane magnetic field angle $\theta$ for total field magnitudes of 9 T (a) and 3 T (b). A sharp increase in $\gamma$ is observed for the high field regime when the magnetic field is in-plane. (c) $\gamma$ is shown versus the in-plane angle $\phi$ for $H_{\parallel}=7.8$ T and $H_{\perp}=4.5$ T (red), $-4.5$ T (black), and 0 T (blue). (d) Similar magnetic field scheme like (c) for the first-harmonic measurements. Note the mirror symmetry breaking for the red and black curves. (e) Sketch of the magnetic field configurations employed in the measurements presented in this figure. Temperatures and sheet resistances for these figures slightly vary due to different cool-downs: (a) $T=3$ K and $R_s$=380 $\Omega$, (b) $T=1.8$ K and $R_s=412$ $\Omega$, (c)-(d) $T=1.9$ K and $R_s=407$ $\Omega$.}
    \label{fig:OutP}
\end{figure*}

The sharp increase of $R_{2\omega}$ shown in Fig. \ref{fig:InP} is extremely sensitive to the application of an out-of-plane magnetic field. This can be seen in Fig. \ref{fig:OutP}(a), where a magnetic field of 9 T ($> H_{cr}$) is rotated out of the plane of the interface, perpendicular to the current (see top image in Fig. \ref{fig:OutP}(e) for clarity). $\gamma$ rapidly decays with the introduction of a small out-of-plane contribution. Conversely, Fig. \ref{fig:OutP}(b) shows the low-field behavior ($H<H_{cr}$), where $\gamma$ is much smaller, and depends linearly on the in-plane magnetic field component. Note that the $V\sim I^2$ dependence of the main non-linear effects studied in this work is confirmed in the supplemental material \cite{sm}, where we present current-dependant second harmonic measurements (Fig. S9) and direct-current versus voltage measurements (Fig. S10) under magnetic fields.

To further investigate the influence of out-of-plane magnetic fields, a constant in-plane field of 7.8 T ($>H_{cr}$) was rotated while applying various out-of-plane fields (see bottom panel of Fig. \ref{fig:OutP}(e)). Fig. \ref{fig:OutP}(c) presents the results for out-of-plane fields of $+4.5$ T (red), $-4.5$ T (black), and 0 T (blue). Although $H>H_{cr}$ in these measurements, the signal with an out-of-plane contribution resembles the behavior observed in the low-field regime, suggesting that an out-of-plane magnetic field mostly counteracts the high field enhancement of $\gamma$.

The same magnetic field measurement scheme used in Fig. \ref{fig:OutP}(c) was employed while measuring the first-harmonic sheet resistance $R_s$ (see Fig. \ref{fig:OutP}(d)). The measurements performed with no out-of-plane field (blue curve) primarily exhibit a two-fold effect as a function of $\phi$, accompanied by four-fold and six-fold contributions (see Fig. S4 \cite{sm}), similar to those previously reported for the (111) LaAlO$_3$/SrTiO$_3$ interface \cite{rout2017six}. However, upon the introduction of an out-of-plane magnetic field (red/black curves), an unusual anisotropy emerges. This anisotropic component breaks the mirror symmetry of the magnetic field with respect to the interface, $R(H_\parallel,H_\perp)\neq R(H_\parallel,-H_\perp)$. Interestingly, for a particular in-plane angle, this mirror symmetry is preserved. This angle (defined as $\phi_0$) exhibits a three-fold dependence on the crystal axis along which current is driven (as seen in Fig. S3 \cite{sm}). While the understanding of this effect lies beyond the scope of this work, its three-fold nature and dependence on the out-of-plane magnetic field suggests that it is a consequence of the three-fold out-of-plane tilt of the spin texture observed for the surface of (111) SrTiO$_3$ \cite{he2018observation}.

\begin{figure}
\centering
\includegraphics[width=\linewidth]{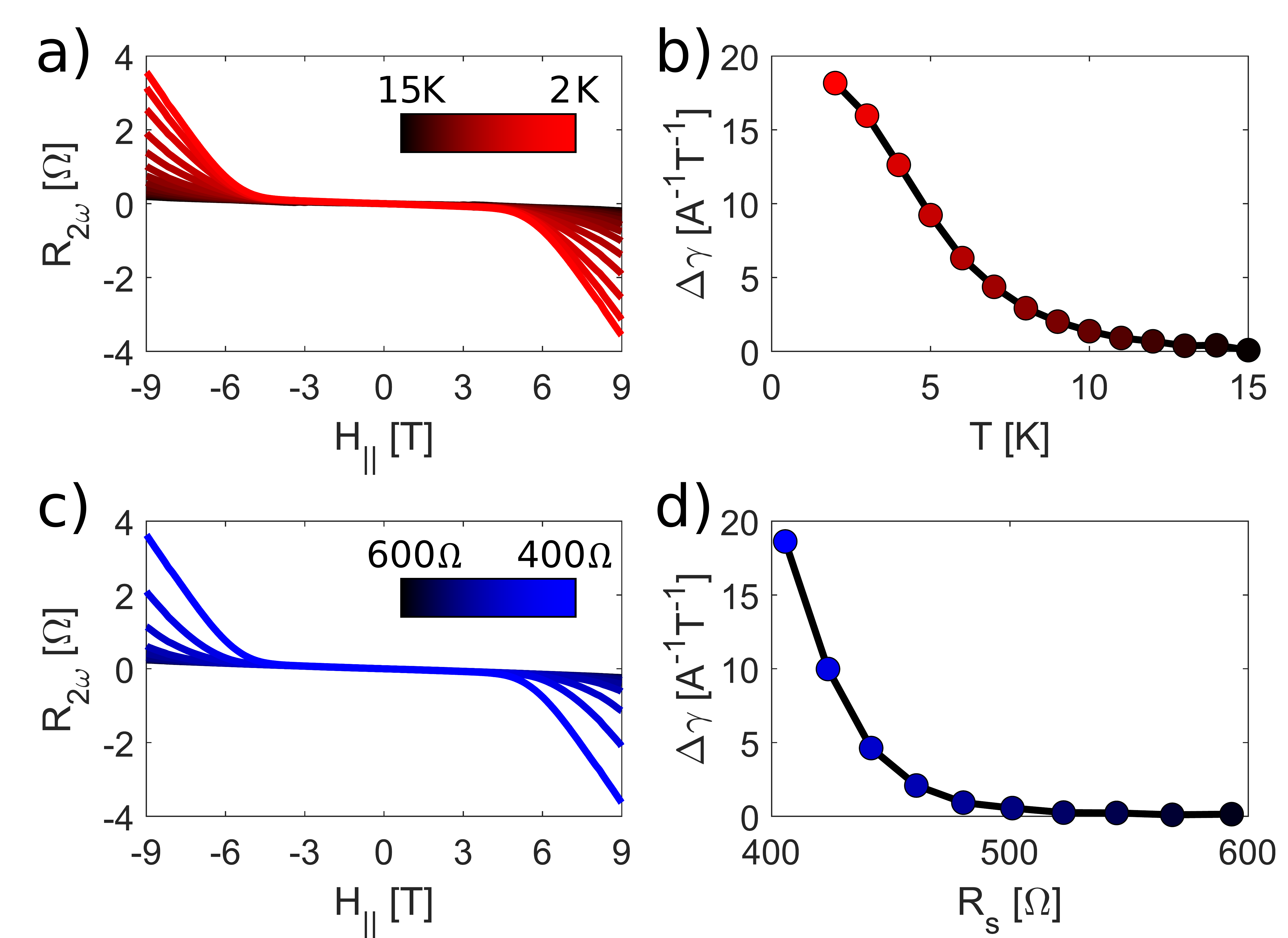}
    \caption{(a) Second-harmonic resistance versus transverse magnetic field at different temperatures. (b) $\gamma$ at 9 T after subtracting the low-field contribution, versus temperature. (c,d) Similar to (a,b) but as a function of gate-voltage represented by the sheet resistance ($R_s$). The effect quickly decays at more negative gates-voltages (higher resistances). Temperature-dependent measurements were conducted for the lowest $R_s$, gate-dependent measurements were conducted at 2 K.} 
\label{fig:TandVg}
\end{figure}

The high-field enhancement of $R_{2\omega}$ decays with temperature, almost vanishing at $\approx10$ K. This behavior can be observed in Fig. \ref{fig:TandVg}(a), where a transverse magnetic field is applied at various temperatures. To quantify the effect of temperature, the $\gamma$ factor at 9 T is plotted in Fig. \ref{fig:TandVg}(b) for various temperatures after subtracting the small low-field contribution, extrapolated to 9 T (the result is defined as $\Delta\gamma$). As shown in Fig. S5 (b) \cite{sm}, the critical field $H_{cr}$ does not change within this temperature range. A similar analysis was performed as a function of gate-voltage and is shown in Fig. \ref{fig:TandVg}(c)-(d). A strong decay of the effect is observed as the gate-voltage becomes more negative. In contrast to the effect of temperature, the critical field $H_{cr}$ changes with the gate-voltage (see Fig. S5 (a) \cite{sm}), reflecting a change of the chemical potential and the SOI magnitude.

To elucidate the critical behavior observed in the second-harmonic transport, we propose a minimal model that can capture the essential features observed in experiment. Our model assumes a two-dimensional electron system with a simple parabolic dispersion and a Rashba-type SOI in the presence of a Zeeman magnetic field:
\begin{equation}\label{eq:H with SOI and By}
\mathcal{H}=\frac{k^2}{2m}+\lambda(k_x\sigma_y-k_y\sigma_x)-B_y\sigma_y-B_z\sigma_z,
\end{equation}
where $\lambda$ represents the magnitude of the SOI, $k_{x/y}$ denote the momentum components which are parallel/perpendicular to the current, $\sigma_{x/y/z}$ are the Pauli matrices, and $B_y/B_z$ are the transverse/out-of-plane magnetic field components. Throughout this work, $B$ is used for magnetic fields in the calculations in units of energy assuming $g$=2, while $H$ is used for the magnetic fields applied in the experiment. Note that we assume that only a single electronic band is occupied. In the supplemental material \cite{sm}, we address a scenario where a second electronic band is occupied, but is very dilute so its effect is unnoticeable in the Hall effect presented in Fig. \ref{fig:SCandHall} (a). We show that such a second-band with reasonable physical parameters cannot account for our experimental observations.

Previous calculations by Ideue \textit{et al.} \cite{ideue2017bulk} have shown that for such a system, the geometry of the Fermi contour changes at critical values of the magnetic field or chemical potential, leading to a change in $\gamma$. To map this scenario onto the (111) LaTiO$_3$/SrTiO$_3$ interface studied here, we estimate the relevant parameters of the system. The spin-orbit energy of SrTiO$_3$ based interfaces is estimated at a few meV \cite{shalom2010tuning,rout2017link,lesne2023designing} while the electron density of a few $10^{13}[\frac{1}{\mathrm{cm^2}}]$ results in a chemical potential of roughly 100meV. To ensure that the estimation of the spin-orbit energy is reasonable, we perform standard magnetoresistance measurements at low temperatures (Fig. S6 \cite{sm}). The relatively small spin-splitting effect means that, for $B = 0$, the spin-split energy bands cross at a single point below the Fermi energy. This crossing point is coined as ``the Dirac point'' due to the linear dispersion in its vicinity. Once a transverse magnetic field $B_y$ is applied, the energy of the Dirac point increases, reaching the Fermi energy at the critical field $B_{cr}=\sqrt{2m\lambda^2\mu}$. Within the relevant parameter regime, the calculated $B_{cr}$ falls reasonably close to the critical field observed in experiment.
\\
\begin{figure}
\includegraphics[width=\linewidth]{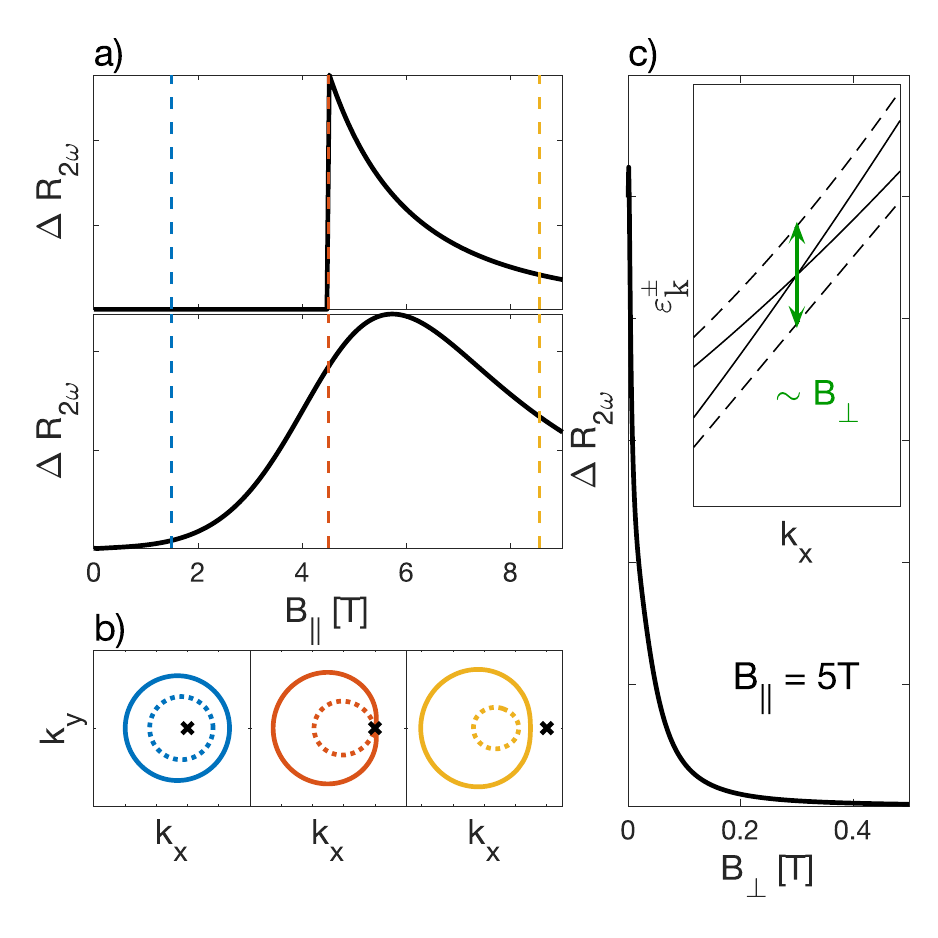}
  \caption{Minimal model calculations. (a) Calculated second-harmonic resistance as function of transverse magnetic field with no out-of-plane field (top panel) and a small out-of-plane field (bottom panel). (b) Fermi contours for $B_{\parallel}<B_{cr},B_{\parallel}=B_{cr},B_{\parallel}>B_{cr}$. For every scenario, the location of the Dirac point is marked. (c) Second-harmonic resistance vs. the perpendicular magnetic field with constant high transverse magnetic field. Inset: Illustration of band gap opening with perpendicular magnetic field.}
    \label{fig:Theory}
\end{figure}

As we now show, the exclusion of the Dirac point from the Fermi contour leads to a dramatic increase in $R_{2\omega}$, similar to our experimental observation. The second-harmonic resistance is proportional to $\sigma_2/\sigma_1^2$, where $J^{(i)} = \sigma_i E^i$. By expanding the Boltzmann equation to a second order in the electric field within the single relaxation time approximation \cite{sm}, we obtain the second order current:
\begin{equation}\label{eq:J2}
J^{(2)}=-e^3\tau^2E^2\sum_{r=\pm}\int\frac{\mathrm{d}k_x\mathrm{d}k_y}{(2\pi)^2}\frac{\partial^3\varepsilon^r_k}{\partial k_x^3}f_0^r,
\end{equation}
where $\tau$ is the scattering time, $E$ is the electric field applied in the $\hat{x}$ direction, $f_0^{\pm}$ are the Fermi-Dirac distributions of the bands, and $\varepsilon^r_k$ are the corresponding energies. Considering first the case of $B_z=0$, the energy bands are given by:
\begin{equation}\label{eq:E(k)}
\varepsilon^\pm_k=\frac{k^2}{2m}\pm \sqrt{(\lambda k_x-B_y)^2+\lambda ^2k_y^2}.
\end{equation}
While the third derivative of the parabolic term $k^2/2m$  in Eq. \eqref{eq:E(k)} vanishes, the Dirac point acts as a source for higher derivatives, and yields a non-vanishing contribution to the integrands in Eq. \eqref{eq:J2}. In section 1 of the supplemental material \cite{sm} we show that for $B_y < B_{cr}$, the sum of third derivatives of the bands is antisymmetric under a $\pi$ rotation around the Dirac point, such that $J^{(2)} = 0$. This symmetry is broken for $B_y > B_{cr}$, where the Dirac point is excluded from both Fermi contours, and $J^{(2)}$ assumes a finite value. Utilizing the above expressions, we calculate $R_{2\omega}$ as function of $B_y$. The result, presented in the top panel of Fig. \ref{fig:Theory}(a), shows that the second harmonic signal is zero for low fields ($B_y < B_{cr}$), significantly increases once the Fermi contours intersect at $B_y = B_{cr}$, and finally decreases for $B_y > B_{cr}$. The position of this Dirac point with respect to the Fermi contours is illustrated in Fig. \ref{fig:Theory}(b) for several values of the transverse magnetic field. We note that the spin-texture of the Fermi contours also changes at the critical field --- for the low-field regime, the spin texture completes a $2\pi$ rotation when going along the Fermi contours while it does not above the critical field \cite{sm}. While this Dirac point mechanism can explain the critical magnetic field observed in experiment, the strong decay of $\gamma$ with temperature and gate-voltage shown in Fig. \ref{fig:TandVg}(b), (d) requires an analysis beyond the single relaxation time approximation used here \cite{sm}.

The presence of an energy gap between the spin-split bands can significantly change the ideal result presented in Fig. \ref{fig:Theory}(a) (top panel). In the bottom panel of Fig. \ref{fig:Theory}(a), we show the calculated $R_{2\omega}$ as a function of the transverse magnetic field $B_y$ when a small gap is opened by a perpendicular magnetic field, equivalent to $B_z=10$ mT. This gap adds a new asymmetry to the energy bands, leading to a small non-zero coefficient below the critical field and a much more gradual increase of $R_{2\omega}$ at the critical field. In Fig. \ref{fig:Theory}(c), we show the calculated $R_{2\omega}$ in a constant transverse magnetic field ($B_y>B_{cr}$) while a perpendicular magnetic field $B_z$ is applied. The second-harmonic signal quickly decays, similar to the experimental result shown in Fig. \ref{fig:OutP}. This decay reflects the fact that the gap smears-out the Dirac point, as seen in Fig. \ref{fig:Theory}(c), which substantially decreases the magnitude of the third derivative.

It is interesting to compare the (111) LaTiO$_3$/SrTiO$_3$ interface studied here to the more extensively studied LaAlO$_3$/SrTiO$_3$ interface. While for both (001) and (111) LaAlO$_3$/SrTiO$_3$ interfaces, the overdoped regime of superconductivity is accompanied by a clear two-band Hall signal \cite{joshua2012universal,maniv2015strong,khanna2019symmetry}. Similarly, certain magneto-transport effects, such as the anomalous Hall effect induced by an in-plane magnetic field \cite{shalom2010tuning,joshua2013gate}, and anisotropic magnetoresistance \cite{joshua2013gate,rout2017six}, manifest in LaAlO$_3$/SrTiO$_3$ interfaces only when a two-band Hall behavior is present. Nevertheless, in the case of the studied interface, these phenomena are observed with a single occupied energy band (refer to Fig. \ref{fig:SCandHall} for overdoped superconductivity, Fig. S8 for in-plane anomalous Hall, and Fig. S4 for magnetoresistance measurements \cite{sm}). This distinction warrants further exploration in future research. It is important to note that, within the context of this study, these magneto-transport effects are evident at low temperatures, high gate voltages, and under the influence of a strong in-plane magnetic field, conditions associated with a large $\gamma$. As noted in previous works \cite{khanna2019symmetry,maniv2015strong,joshua2012universal}, the Rashba SOI only adds a small correction to the band structure. However, as we show in this work, the manipulation of the SO induced Dirac point changes the symmetry of the energy bands, which has a substantial effect on the non-linear transport, and possibly also the linear magneto-transport.

In summary, we measure the new (111) LaTiO$_3$/SrTiO$_3$ interface. We observe superconductivity in this interface along with one-band-type Hall-signal. This behavior is observed even when T$_c$ is decreased upon adding charge carriers (overdoped regime), in contrast to the well studied LaAlO$_3$/SrTiO$_3$ interfaces. Our main finding is a sharp increase of the magnitude of the non-linear transport of the (111) LaTiO$_3$/SrTiO$_3$ interface at a critical magnetic field $H_{cr}$. By employing a minimal model incorporating SOI at a generic interface, we show that $H_{cr}$ can be understood as the field at which the spin-orbit-induced Dirac point is excluded from the Fermi contour. According to our model, beyond this field, the symmetry and the spin-texture of the Fermi contours change simultaneously. We further show that a minuscule out-of-plane magnetic field opens a gap between the spin-split bands, which significantly reduces the second-harmonic effect, as observed in our experiment. We suggest that this mechanism sheds light on some of the unusual magneto-transport effects observed in previous experiments on oxide interfaces. 
\subsection*{Acknowledgements}
We thank Tobias Holder, Roni Ilan, and Shay Sandik for useful discussions. Asaf Yagoda and Elad Mileikowsky assisted with the transport setup. This work was partially supported (Y.D.) By the Israeli Science Foundation under grant 476/22  and by the PAZI foundation under grant 326-1/20. M.G. was supported by the Israel Science Foundation (ISF) and the Directorate for Defense Research and Development (DDR\&D) Grant No. 3427/21, and by the US-Israel Binational Science Foundation (BSF) Grant No. 2020072. A.B. is supported by the Adams Fellowship Program of the Israel Academy of Sciences and Humanities.
\bibliographystyle{apsrev4-2}
\bibliography{references}

\begin{thebibliography}{28}%
\makeatletter
\providecommand \@ifxundefined [1]{%
 \@ifx{#1\undefined}
}%
\providecommand \@ifnum [1]{%
 \ifnum #1\expandafter \@firstoftwo
 \else \expandafter \@secondoftwo
 \fi
}%
\providecommand \@ifx [1]{%
 \ifx #1\expandafter \@firstoftwo
 \else \expandafter \@secondoftwo
 \fi
}%
\providecommand \natexlab [1]{#1}%
\providecommand \enquote  [1]{``#1''}%
\providecommand \bibnamefont  [1]{#1}%
\providecommand \bibfnamefont [1]{#1}%
\providecommand \citenamefont [1]{#1}%
\providecommand \href@noop [0]{\@secondoftwo}%
\providecommand \href [0]{\begingroup \@sanitize@url \@href}%
\providecommand \@href[1]{\@@startlink{#1}\@@href}%
\providecommand \@@href[1]{\endgroup#1\@@endlink}%
\providecommand \@sanitize@url [0]{\catcode `\\12\catcode `\$12\catcode `\&12\catcode `\#12\catcode `\^12\catcode `\_12\catcode `\%12\relax}%
\providecommand \@@startlink[1]{}%
\providecommand \@@endlink[0]{}%
\providecommand \url  [0]{\begingroup\@sanitize@url \@url }%
\providecommand \@url [1]{\endgroup\@href {#1}{\urlprefix }}%
\providecommand \urlprefix  [0]{URL }%
\providecommand \Eprint [0]{\href }%
\providecommand \doibase [0]{https://doi.org/}%
\providecommand \selectlanguage [0]{\@gobble}%
\providecommand \bibinfo  [0]{\@secondoftwo}%
\providecommand \bibfield  [0]{\@secondoftwo}%
\providecommand \translation [1]{[#1]}%
\providecommand \BibitemOpen [0]{}%
\providecommand \bibitemStop [0]{}%
\providecommand \bibitemNoStop [0]{.\EOS\space}%
\providecommand \EOS [0]{\spacefactor3000\relax}%
\providecommand \BibitemShut  [1]{\csname bibitem#1\endcsname}%
\let\auto@bib@innerbib\@empty
\bibitem [{\citenamefont {Rashba}(1960)}]{rashba1960properties}%
  \BibitemOpen
  \bibfield  {author} {\bibinfo {author} {\bibfnamefont {E.}~\bibnamefont {Rashba}},\ }\href {https://cir.nii.ac.jp/crid/1571698600346713472} {\bibfield  {journal} {\bibinfo  {journal} {Sov. Phys.-Solid State}\ }\textbf {\bibinfo {volume} {2}},\ \bibinfo {pages} {1109} (\bibinfo {year} {1960})}\BibitemShut {NoStop}%
\bibitem [{\citenamefont {Ishizaka}\ \emph {et~al.}(2011)\citenamefont {Ishizaka}, \citenamefont {Bahramy}, \citenamefont {Murakawa}, \citenamefont {Sakano}, \citenamefont {Shimojima}, \citenamefont {Sonobe}, \citenamefont {Koizumi}, \citenamefont {Shin}, \citenamefont {Miyahara}, \citenamefont {Kimura}, \citenamefont {Miyamoto}, \citenamefont {Okuda}, \citenamefont {Namatame}, \citenamefont {Taniguchi}, \citenamefont {Arita}, \citenamefont {Nagaosa}, \citenamefont {Kobayashi}, \citenamefont {Murakami}, \citenamefont {Kumai}, \citenamefont {Kaneko}, \citenamefont {Onose},\ and\ \citenamefont {Tokura}}]{ishizaka2011giant}%
  \BibitemOpen
  \bibfield  {author} {\bibinfo {author} {\bibfnamefont {K.}~\bibnamefont {Ishizaka}}, \bibinfo {author} {\bibfnamefont {M.~S.}\ \bibnamefont {Bahramy}}, \bibinfo {author} {\bibfnamefont {H.}~\bibnamefont {Murakawa}}, \bibinfo {author} {\bibfnamefont {M.}~\bibnamefont {Sakano}}, \bibinfo {author} {\bibfnamefont {T.}~\bibnamefont {Shimojima}}, \bibinfo {author} {\bibfnamefont {T.}~\bibnamefont {Sonobe}}, \bibinfo {author} {\bibfnamefont {K.}~\bibnamefont {Koizumi}}, \bibinfo {author} {\bibfnamefont {S.}~\bibnamefont {Shin}}, \bibinfo {author} {\bibfnamefont {H.}~\bibnamefont {Miyahara}}, \bibinfo {author} {\bibfnamefont {A.}~\bibnamefont {Kimura}}, \bibinfo {author} {\bibfnamefont {K.}~\bibnamefont {Miyamoto}}, \bibinfo {author} {\bibfnamefont {T.}~\bibnamefont {Okuda}}, \bibinfo {author} {\bibfnamefont {H.}~\bibnamefont {Namatame}}, \bibinfo {author} {\bibfnamefont {M.}~\bibnamefont {Taniguchi}}, \bibinfo {author} {\bibfnamefont {R.}~\bibnamefont {Arita}}, \bibinfo {author} {\bibfnamefont {N.}~\bibnamefont
  {Nagaosa}}, \bibinfo {author} {\bibfnamefont {K.}~\bibnamefont {Kobayashi}}, \bibinfo {author} {\bibfnamefont {Y.}~\bibnamefont {Murakami}}, \bibinfo {author} {\bibfnamefont {R.}~\bibnamefont {Kumai}}, \bibinfo {author} {\bibfnamefont {Y.}~\bibnamefont {Kaneko}}, \bibinfo {author} {\bibfnamefont {Y.}~\bibnamefont {Onose}},\ and\ \bibinfo {author} {\bibfnamefont {Y.}~\bibnamefont {Tokura}},\ }\href {https://doi.org/10.1038/nmat3051} {\bibfield  {journal} {\bibinfo  {journal} {Nature Materials}\ }\textbf {\bibinfo {volume} {10}},\ \bibinfo {pages} {521} (\bibinfo {year} {2011})}\BibitemShut {NoStop}%
\bibitem [{\citenamefont {Ideue}\ \emph {et~al.}(2017)\citenamefont {Ideue}, \citenamefont {Hamamoto}, \citenamefont {Koshikawa}, \citenamefont {Ezawa}, \citenamefont {Shimizu}, \citenamefont {Kaneko}, \citenamefont {Tokura}, \citenamefont {Nagaosa},\ and\ \citenamefont {Iwasa}}]{ideue2017bulk}%
  \BibitemOpen
  \bibfield  {author} {\bibinfo {author} {\bibfnamefont {T.}~\bibnamefont {Ideue}}, \bibinfo {author} {\bibfnamefont {K.}~\bibnamefont {Hamamoto}}, \bibinfo {author} {\bibfnamefont {S.}~\bibnamefont {Koshikawa}}, \bibinfo {author} {\bibfnamefont {M.}~\bibnamefont {Ezawa}}, \bibinfo {author} {\bibfnamefont {S.}~\bibnamefont {Shimizu}}, \bibinfo {author} {\bibfnamefont {Y.}~\bibnamefont {Kaneko}}, \bibinfo {author} {\bibfnamefont {Y.}~\bibnamefont {Tokura}}, \bibinfo {author} {\bibfnamefont {N.}~\bibnamefont {Nagaosa}},\ and\ \bibinfo {author} {\bibfnamefont {Y.}~\bibnamefont {Iwasa}},\ }\href {https://doi.org/10.1038/nphys4056} {\bibfield  {journal} {\bibinfo  {journal} {Nature Physics}\ }\textbf {\bibinfo {volume} {13}},\ \bibinfo {pages} {578} (\bibinfo {year} {2017})}\BibitemShut {NoStop}%
\bibitem [{\citenamefont {Santander-Syro}\ \emph {et~al.}(2014)\citenamefont {Santander-Syro}, \citenamefont {Fortuna}, \citenamefont {Bareille}, \citenamefont {R{\"o}del}, \citenamefont {Landolt}, \citenamefont {Plumb}, \citenamefont {Dil},\ and\ \citenamefont {Radovi{\'{c}}}}]{santander2014giant}%
  \BibitemOpen
  \bibfield  {author} {\bibinfo {author} {\bibfnamefont {A.~F.}\ \bibnamefont {Santander-Syro}}, \bibinfo {author} {\bibfnamefont {F.}~\bibnamefont {Fortuna}}, \bibinfo {author} {\bibfnamefont {C.}~\bibnamefont {Bareille}}, \bibinfo {author} {\bibfnamefont {T.~C.}\ \bibnamefont {R{\"o}del}}, \bibinfo {author} {\bibfnamefont {G.}~\bibnamefont {Landolt}}, \bibinfo {author} {\bibfnamefont {N.~C.}\ \bibnamefont {Plumb}}, \bibinfo {author} {\bibfnamefont {J.~H.}\ \bibnamefont {Dil}},\ and\ \bibinfo {author} {\bibfnamefont {M.}~\bibnamefont {Radovi{\'{c}}}},\ }\href {https://doi.org/10.1038/nmat4107} {\bibfield  {journal} {\bibinfo  {journal} {Nature Materials}\ }\textbf {\bibinfo {volume} {13}},\ \bibinfo {pages} {1085} (\bibinfo {year} {2014})}\BibitemShut {NoStop}%
\bibitem [{\citenamefont {LaShell}\ \emph {et~al.}(1996)\citenamefont {LaShell}, \citenamefont {McDougall},\ and\ \citenamefont {Jensen}}]{lashell1996spin}%
  \BibitemOpen
  \bibfield  {author} {\bibinfo {author} {\bibfnamefont {S.}~\bibnamefont {LaShell}}, \bibinfo {author} {\bibfnamefont {B.~A.}\ \bibnamefont {McDougall}},\ and\ \bibinfo {author} {\bibfnamefont {E.}~\bibnamefont {Jensen}},\ }\href {https://doi.org/10.1103/PhysRevLett.77.3419} {\bibfield  {journal} {\bibinfo  {journal} {Phys. Rev. Lett.}\ }\textbf {\bibinfo {volume} {77}},\ \bibinfo {pages} {3419} (\bibinfo {year} {1996})}\BibitemShut {NoStop}%
\bibitem [{\citenamefont {Manchon}\ \emph {et~al.}(2015)\citenamefont {Manchon}, \citenamefont {Koo}, \citenamefont {Nitta}, \citenamefont {Frolov},\ and\ \citenamefont {Duine}}]{manchon2015new}%
  \BibitemOpen
  \bibfield  {author} {\bibinfo {author} {\bibfnamefont {A.}~\bibnamefont {Manchon}}, \bibinfo {author} {\bibfnamefont {H.~C.}\ \bibnamefont {Koo}}, \bibinfo {author} {\bibfnamefont {J.}~\bibnamefont {Nitta}}, \bibinfo {author} {\bibfnamefont {S.~M.}\ \bibnamefont {Frolov}},\ and\ \bibinfo {author} {\bibfnamefont {R.~A.}\ \bibnamefont {Duine}},\ }\href {https://doi.org/10.1038/nmat4360} {\bibfield  {journal} {\bibinfo  {journal} {Nature Materials}\ }\textbf {\bibinfo {volume} {14}},\ \bibinfo {pages} {871} (\bibinfo {year} {2015})}\BibitemShut {NoStop}%
\bibitem [{\citenamefont {Premasiri}\ and\ \citenamefont {Gao}(2019)}]{premasiri2019tuning}%
  \BibitemOpen
  \bibfield  {author} {\bibinfo {author} {\bibfnamefont {K.}~\bibnamefont {Premasiri}}\ and\ \bibinfo {author} {\bibfnamefont {X.~P.}\ \bibnamefont {Gao}},\ }\href {https://doi.org/10.1088/1361-648X/ab04c7} {\bibfield  {journal} {\bibinfo  {journal} {Journal of Physics: Condensed Matter}\ }\textbf {\bibinfo {volume} {31}},\ \bibinfo {pages} {193001} (\bibinfo {year} {2019})}\BibitemShut {NoStop}%
\bibitem [{\citenamefont {Rakhmilevitch}\ \emph {et~al.}(2010)\citenamefont {Rakhmilevitch}, \citenamefont {Ben~Shalom}, \citenamefont {Eshkol}, \citenamefont {Tsukernik}, \citenamefont {Palevski},\ and\ \citenamefont {Dagan}}]{rakhmilevitch2010phase}%
  \BibitemOpen
  \bibfield  {author} {\bibinfo {author} {\bibfnamefont {D.}~\bibnamefont {Rakhmilevitch}}, \bibinfo {author} {\bibfnamefont {M.}~\bibnamefont {Ben~Shalom}}, \bibinfo {author} {\bibfnamefont {M.}~\bibnamefont {Eshkol}}, \bibinfo {author} {\bibfnamefont {A.}~\bibnamefont {Tsukernik}}, \bibinfo {author} {\bibfnamefont {A.}~\bibnamefont {Palevski}},\ and\ \bibinfo {author} {\bibfnamefont {Y.}~\bibnamefont {Dagan}},\ }\href {https://doi.org/10.1103/PhysRevB.82.235119} {\bibfield  {journal} {\bibinfo  {journal} {Phys. Rev. B}\ }\textbf {\bibinfo {volume} {82}},\ \bibinfo {pages} {235119} (\bibinfo {year} {2010})}\BibitemShut {NoStop}%
\bibitem [{\citenamefont {Ben~Shalom}\ \emph {et~al.}(2010)\citenamefont {Ben~Shalom}, \citenamefont {Sachs}, \citenamefont {Rakhmilevitch}, \citenamefont {Palevski},\ and\ \citenamefont {Dagan}}]{shalom2010tuning}%
  \BibitemOpen
  \bibfield  {author} {\bibinfo {author} {\bibfnamefont {M.}~\bibnamefont {Ben~Shalom}}, \bibinfo {author} {\bibfnamefont {M.}~\bibnamefont {Sachs}}, \bibinfo {author} {\bibfnamefont {D.}~\bibnamefont {Rakhmilevitch}}, \bibinfo {author} {\bibfnamefont {A.}~\bibnamefont {Palevski}},\ and\ \bibinfo {author} {\bibfnamefont {Y.}~\bibnamefont {Dagan}},\ }\href {https://doi.org/10.1103/PhysRevLett.104.126802} {\bibfield  {journal} {\bibinfo  {journal} {Phys. Rev. Lett.}\ }\textbf {\bibinfo {volume} {104}},\ \bibinfo {pages} {126802} (\bibinfo {year} {2010})}\BibitemShut {NoStop}%
\bibitem [{\citenamefont {Rout}\ \emph {et~al.}(2017{\natexlab{a}})\citenamefont {Rout}, \citenamefont {Maniv},\ and\ \citenamefont {Dagan}}]{rout2017link}%
  \BibitemOpen
  \bibfield  {author} {\bibinfo {author} {\bibfnamefont {P.~K.}\ \bibnamefont {Rout}}, \bibinfo {author} {\bibfnamefont {E.}~\bibnamefont {Maniv}},\ and\ \bibinfo {author} {\bibfnamefont {Y.}~\bibnamefont {Dagan}},\ }\href {https://doi.org/10.1103/PhysRevLett.119.237002} {\bibfield  {journal} {\bibinfo  {journal} {Phys. Rev. Lett.}\ }\textbf {\bibinfo {volume} {119}},\ \bibinfo {pages} {237002} (\bibinfo {year} {2017}{\natexlab{a}})}\BibitemShut {NoStop}%
\bibitem [{\citenamefont {Lesne}\ \emph {et~al.}(2023)\citenamefont {Lesne}, \citenamefont {Sa{\v{g}}lam}, \citenamefont {Battilomo}, \citenamefont {Mercaldo}, \citenamefont {van Thiel}, \citenamefont {Filippozzi}, \citenamefont {Noce}, \citenamefont {Cuoco}, \citenamefont {Steele}, \citenamefont {Ortix},\ and\ \citenamefont {Caviglia}}]{lesne2023designing}%
  \BibitemOpen
  \bibfield  {author} {\bibinfo {author} {\bibfnamefont {E.}~\bibnamefont {Lesne}}, \bibinfo {author} {\bibfnamefont {Y.~G.}\ \bibnamefont {Sa{\v{g}}lam}}, \bibinfo {author} {\bibfnamefont {R.}~\bibnamefont {Battilomo}}, \bibinfo {author} {\bibfnamefont {M.~T.}\ \bibnamefont {Mercaldo}}, \bibinfo {author} {\bibfnamefont {T.~C.}\ \bibnamefont {van Thiel}}, \bibinfo {author} {\bibfnamefont {U.}~\bibnamefont {Filippozzi}}, \bibinfo {author} {\bibfnamefont {C.}~\bibnamefont {Noce}}, \bibinfo {author} {\bibfnamefont {M.}~\bibnamefont {Cuoco}}, \bibinfo {author} {\bibfnamefont {G.~A.}\ \bibnamefont {Steele}}, \bibinfo {author} {\bibfnamefont {C.}~\bibnamefont {Ortix}},\ and\ \bibinfo {author} {\bibfnamefont {A.~D.}\ \bibnamefont {Caviglia}},\ }\href {https://doi.org/10.1038/s41563-023-01498-0} {\bibfield  {journal} {\bibinfo  {journal} {Nature Materials}\ }\textbf {\bibinfo {volume} {22}},\ \bibinfo {pages} {576} (\bibinfo {year} {2023})}\BibitemShut {NoStop}%
\bibitem [{\citenamefont {He}\ \emph {et~al.}(2018)\citenamefont {He}, \citenamefont {Walker}, \citenamefont {Zhang}, \citenamefont {Bruno}, \citenamefont {Bahramy}, \citenamefont {Lee}, \citenamefont {Ramaswamy}, \citenamefont {Cai}, \citenamefont {Heinonen}, \citenamefont {Vignale}, \citenamefont {Baumberger},\ and\ \citenamefont {Yang}}]{he2018observation}%
  \BibitemOpen
  \bibfield  {author} {\bibinfo {author} {\bibfnamefont {P.}~\bibnamefont {He}}, \bibinfo {author} {\bibfnamefont {S.~M.}\ \bibnamefont {Walker}}, \bibinfo {author} {\bibfnamefont {S.~S.-L.}\ \bibnamefont {Zhang}}, \bibinfo {author} {\bibfnamefont {F.~Y.}\ \bibnamefont {Bruno}}, \bibinfo {author} {\bibfnamefont {M.~S.}\ \bibnamefont {Bahramy}}, \bibinfo {author} {\bibfnamefont {J.~M.}\ \bibnamefont {Lee}}, \bibinfo {author} {\bibfnamefont {R.}~\bibnamefont {Ramaswamy}}, \bibinfo {author} {\bibfnamefont {K.}~\bibnamefont {Cai}}, \bibinfo {author} {\bibfnamefont {O.}~\bibnamefont {Heinonen}}, \bibinfo {author} {\bibfnamefont {G.}~\bibnamefont {Vignale}}, \bibinfo {author} {\bibfnamefont {F.}~\bibnamefont {Baumberger}},\ and\ \bibinfo {author} {\bibfnamefont {H.}~\bibnamefont {Yang}},\ }\href {https://doi.org/10.1103/PhysRevLett.120.266802} {\bibfield  {journal} {\bibinfo  {journal} {Phys. Rev. Lett.}\ }\textbf {\bibinfo {volume} {120}},\ \bibinfo {pages} {266802} (\bibinfo {year} {2018})}\BibitemShut {NoStop}%
\bibitem [{\citenamefont {Rout}\ \emph {et~al.}(2017{\natexlab{b}})\citenamefont {Rout}, \citenamefont {Agireen}, \citenamefont {Maniv}, \citenamefont {Goldstein},\ and\ \citenamefont {Dagan}}]{rout2017six}%
  \BibitemOpen
  \bibfield  {author} {\bibinfo {author} {\bibfnamefont {P.~K.}\ \bibnamefont {Rout}}, \bibinfo {author} {\bibfnamefont {I.}~\bibnamefont {Agireen}}, \bibinfo {author} {\bibfnamefont {E.}~\bibnamefont {Maniv}}, \bibinfo {author} {\bibfnamefont {M.}~\bibnamefont {Goldstein}},\ and\ \bibinfo {author} {\bibfnamefont {Y.}~\bibnamefont {Dagan}},\ }\href {https://doi.org/10.1103/PhysRevB.95.241107} {\bibfield  {journal} {\bibinfo  {journal} {Phys. Rev. B}\ }\textbf {\bibinfo {volume} {95}},\ \bibinfo {pages} {241107} (\bibinfo {year} {2017}{\natexlab{b}})}\BibitemShut {NoStop}%
\bibitem [{\citenamefont {Isobe}\ \emph {et~al.}(2020)\citenamefont {Isobe}, \citenamefont {Xu},\ and\ \citenamefont {Fu}}]{isobe2020high}%
  \BibitemOpen
  \bibfield  {author} {\bibinfo {author} {\bibfnamefont {H.}~\bibnamefont {Isobe}}, \bibinfo {author} {\bibfnamefont {S.-Y.}\ \bibnamefont {Xu}},\ and\ \bibinfo {author} {\bibfnamefont {L.}~\bibnamefont {Fu}},\ }\href {https://doi.org/10.1126/sciadv.aay2497} {\bibfield  {journal} {\bibinfo  {journal} {Science Advances}\ }\textbf {\bibinfo {volume} {6}},\ \bibinfo {pages} {eaay2497} (\bibinfo {year} {2020})}\BibitemShut {NoStop}%
\bibitem [{\citenamefont {He}\ \emph {et~al.}(2021)\citenamefont {He}, \citenamefont {Isobe}, \citenamefont {Zhu}, \citenamefont {Hsu}, \citenamefont {Fu},\ and\ \citenamefont {Yang}}]{he2021quantum}%
  \BibitemOpen
  \bibfield  {author} {\bibinfo {author} {\bibfnamefont {P.}~\bibnamefont {He}}, \bibinfo {author} {\bibfnamefont {H.}~\bibnamefont {Isobe}}, \bibinfo {author} {\bibfnamefont {D.}~\bibnamefont {Zhu}}, \bibinfo {author} {\bibfnamefont {C.-H.}\ \bibnamefont {Hsu}}, \bibinfo {author} {\bibfnamefont {L.}~\bibnamefont {Fu}},\ and\ \bibinfo {author} {\bibfnamefont {H.}~\bibnamefont {Yang}},\ }\href {https://doi.org/10.1038/s41467-021-20983-1} {\bibfield  {journal} {\bibinfo  {journal} {Nature Communications}\ }\textbf {\bibinfo {volume} {12}},\ \bibinfo {pages} {698} (\bibinfo {year} {2021})}\BibitemShut {NoStop}%
\bibitem [{\citenamefont {Itahashi}\ \emph {et~al.}(2020)\citenamefont {Itahashi}, \citenamefont {Ideue}, \citenamefont {Saito}, \citenamefont {Shimizu}, \citenamefont {Ouchi}, \citenamefont {Nojima},\ and\ \citenamefont {Iwasa}}]{itahashi2020nonreciprocal}%
  \BibitemOpen
  \bibfield  {author} {\bibinfo {author} {\bibfnamefont {Y.~M.}\ \bibnamefont {Itahashi}}, \bibinfo {author} {\bibfnamefont {T.}~\bibnamefont {Ideue}}, \bibinfo {author} {\bibfnamefont {Y.}~\bibnamefont {Saito}}, \bibinfo {author} {\bibfnamefont {S.}~\bibnamefont {Shimizu}}, \bibinfo {author} {\bibfnamefont {T.}~\bibnamefont {Ouchi}}, \bibinfo {author} {\bibfnamefont {T.}~\bibnamefont {Nojima}},\ and\ \bibinfo {author} {\bibfnamefont {Y.}~\bibnamefont {Iwasa}},\ }\href {https://doi.org/10.1126/sciadv.aay9120} {\bibfield  {journal} {\bibinfo  {journal} {Science Advances}\ }\textbf {\bibinfo {volume} {6}},\ \bibinfo {pages} {eaay9120} (\bibinfo {year} {2020})}\BibitemShut {NoStop}%
\bibitem [{\citenamefont {Itahashi}\ \emph {et~al.}(2022)\citenamefont {Itahashi}, \citenamefont {Ideue}, \citenamefont {Hoshino}, \citenamefont {Goto}, \citenamefont {Namiki}, \citenamefont {Sasagawa},\ and\ \citenamefont {Iwasa}}]{itahashi2022giant}%
  \BibitemOpen
  \bibfield  {author} {\bibinfo {author} {\bibfnamefont {Y.~M.}\ \bibnamefont {Itahashi}}, \bibinfo {author} {\bibfnamefont {T.}~\bibnamefont {Ideue}}, \bibinfo {author} {\bibfnamefont {S.}~\bibnamefont {Hoshino}}, \bibinfo {author} {\bibfnamefont {C.}~\bibnamefont {Goto}}, \bibinfo {author} {\bibfnamefont {H.}~\bibnamefont {Namiki}}, \bibinfo {author} {\bibfnamefont {T.}~\bibnamefont {Sasagawa}},\ and\ \bibinfo {author} {\bibfnamefont {Y.}~\bibnamefont {Iwasa}},\ }\href {https://doi.org/10.1038/s41467-022-29314-4} {\bibfield  {journal} {\bibinfo  {journal} {Nature Communications}\ }\textbf {\bibinfo {volume} {13}},\ \bibinfo {pages} {1659} (\bibinfo {year} {2022})}\BibitemShut {NoStop}%
\bibitem [{\citenamefont {He}\ \emph {et~al.}(2019)\citenamefont {He}, \citenamefont {Hsu}, \citenamefont {Shi}, \citenamefont {Cai}, \citenamefont {Wang}, \citenamefont {Wang}, \citenamefont {Eda}, \citenamefont {Lin}, \citenamefont {Pereira},\ and\ \citenamefont {Yang}}]{he2019nonlinear}%
  \BibitemOpen
  \bibfield  {author} {\bibinfo {author} {\bibfnamefont {P.}~\bibnamefont {He}}, \bibinfo {author} {\bibfnamefont {C.-H.}\ \bibnamefont {Hsu}}, \bibinfo {author} {\bibfnamefont {S.}~\bibnamefont {Shi}}, \bibinfo {author} {\bibfnamefont {K.}~\bibnamefont {Cai}}, \bibinfo {author} {\bibfnamefont {J.}~\bibnamefont {Wang}}, \bibinfo {author} {\bibfnamefont {Q.}~\bibnamefont {Wang}}, \bibinfo {author} {\bibfnamefont {G.}~\bibnamefont {Eda}}, \bibinfo {author} {\bibfnamefont {H.}~\bibnamefont {Lin}}, \bibinfo {author} {\bibfnamefont {V.~M.}\ \bibnamefont {Pereira}},\ and\ \bibinfo {author} {\bibfnamefont {H.}~\bibnamefont {Yang}},\ }\href {https://doi.org/10.1038/s41467-019-09208-8} {\bibfield  {journal} {\bibinfo  {journal} {Nature Communications}\ }\textbf {\bibinfo {volume} {10}},\ \bibinfo {pages} {1290} (\bibinfo {year} {2019})}\BibitemShut {NoStop}%
\bibitem [{\citenamefont {Choe}\ \emph {et~al.}(2019)\citenamefont {Choe}, \citenamefont {Jin}, \citenamefont {Kim}, \citenamefont {Choi}, \citenamefont {Jo}, \citenamefont {Oh}, \citenamefont {Park}, \citenamefont {Jin}, \citenamefont {Koo}, \citenamefont {Min}, \citenamefont {Hong}, \citenamefont {Lee}, \citenamefont {Baek},\ and\ \citenamefont {Yoo}}]{choe2019gate}%
  \BibitemOpen
  \bibfield  {author} {\bibinfo {author} {\bibfnamefont {D.}~\bibnamefont {Choe}}, \bibinfo {author} {\bibfnamefont {M.-J.}\ \bibnamefont {Jin}}, \bibinfo {author} {\bibfnamefont {S.-I.}\ \bibnamefont {Kim}}, \bibinfo {author} {\bibfnamefont {H.-J.}\ \bibnamefont {Choi}}, \bibinfo {author} {\bibfnamefont {J.}~\bibnamefont {Jo}}, \bibinfo {author} {\bibfnamefont {I.}~\bibnamefont {Oh}}, \bibinfo {author} {\bibfnamefont {J.}~\bibnamefont {Park}}, \bibinfo {author} {\bibfnamefont {H.}~\bibnamefont {Jin}}, \bibinfo {author} {\bibfnamefont {H.~C.}\ \bibnamefont {Koo}}, \bibinfo {author} {\bibfnamefont {B.-C.}\ \bibnamefont {Min}}, \bibinfo {author} {\bibfnamefont {S.}~\bibnamefont {Hong}}, \bibinfo {author} {\bibfnamefont {H.-W.}\ \bibnamefont {Lee}}, \bibinfo {author} {\bibfnamefont {S.-H.}\ \bibnamefont {Baek}},\ and\ \bibinfo {author} {\bibfnamefont {J.-W.}\ \bibnamefont {Yoo}},\ }\href {https://doi.org/10.1038/s41467-019-12466-1} {\bibfield  {journal} {\bibinfo  {journal} {Nature Communications}\ }\textbf
  {\bibinfo {volume} {10}},\ \bibinfo {pages} {4510} (\bibinfo {year} {2019})}\BibitemShut {NoStop}%
\bibitem [{\citenamefont {Li}\ \emph {et~al.}(2021)\citenamefont {Li}, \citenamefont {Li}, \citenamefont {Li}, \citenamefont {Fang}, \citenamefont {Yang}, \citenamefont {Wen}, \citenamefont {Zheng}, \citenamefont {Zhang}, \citenamefont {He}, \citenamefont {Manchon}, \citenamefont {Cheng},\ and\ \citenamefont {Zhang}}]{li2021nonreciprocal}%
  \BibitemOpen
  \bibfield  {author} {\bibinfo {author} {\bibfnamefont {Y.}~\bibnamefont {Li}}, \bibinfo {author} {\bibfnamefont {Y.}~\bibnamefont {Li}}, \bibinfo {author} {\bibfnamefont {P.}~\bibnamefont {Li}}, \bibinfo {author} {\bibfnamefont {B.}~\bibnamefont {Fang}}, \bibinfo {author} {\bibfnamefont {X.}~\bibnamefont {Yang}}, \bibinfo {author} {\bibfnamefont {Y.}~\bibnamefont {Wen}}, \bibinfo {author} {\bibfnamefont {D.-x.}\ \bibnamefont {Zheng}}, \bibinfo {author} {\bibfnamefont {C.-h.}\ \bibnamefont {Zhang}}, \bibinfo {author} {\bibfnamefont {X.}~\bibnamefont {He}}, \bibinfo {author} {\bibfnamefont {A.}~\bibnamefont {Manchon}}, \bibinfo {author} {\bibfnamefont {Z.-H.}\ \bibnamefont {Cheng}},\ and\ \bibinfo {author} {\bibfnamefont {X.-x.}\ \bibnamefont {Zhang}},\ }\href {https://doi.org/10.1038/s41467-020-20840-7} {\bibfield  {journal} {\bibinfo  {journal} {Nature Communications}\ }\textbf {\bibinfo {volume} {12}},\ \bibinfo {pages} {540} (\bibinfo {year} {2021})}\BibitemShut {NoStop}%
\bibitem [{\citenamefont {Ohtomo}\ \emph {et~al.}(2002)\citenamefont {Ohtomo}, \citenamefont {Muller}, \citenamefont {Grazul},\ and\ \citenamefont {Hwang}}]{ohtomo2002artificial}%
  \BibitemOpen
  \bibfield  {author} {\bibinfo {author} {\bibfnamefont {A.}~\bibnamefont {Ohtomo}}, \bibinfo {author} {\bibfnamefont {D.~A.}\ \bibnamefont {Muller}}, \bibinfo {author} {\bibfnamefont {J.~L.}\ \bibnamefont {Grazul}},\ and\ \bibinfo {author} {\bibfnamefont {H.~Y.}\ \bibnamefont {Hwang}},\ }\href {https://doi.org/10.1038/nature00977} {\bibfield  {journal} {\bibinfo  {journal} {Nature}\ }\textbf {\bibinfo {volume} {419}},\ \bibinfo {pages} {378} (\bibinfo {year} {2002})}\BibitemShut {NoStop}%
\bibitem [{\citenamefont {Wang}\ \emph {et~al.}(2020)\citenamefont {Wang}, \citenamefont {Liang}, \citenamefont {Meng}, \citenamefont {An}, \citenamefont {Ge}, \citenamefont {Liu}, \citenamefont {Yang},\ and\ \citenamefont {Guo}}]{wang2020anomalous}%
  \BibitemOpen
  \bibfield  {author} {\bibinfo {author} {\bibfnamefont {Y.}~\bibnamefont {Wang}}, \bibinfo {author} {\bibfnamefont {Y.}~\bibnamefont {Liang}}, \bibinfo {author} {\bibfnamefont {M.}~\bibnamefont {Meng}}, \bibinfo {author} {\bibfnamefont {Q.}~\bibnamefont {An}}, \bibinfo {author} {\bibfnamefont {B.}~\bibnamefont {Ge}}, \bibinfo {author} {\bibfnamefont {M.}~\bibnamefont {Liu}}, \bibinfo {author} {\bibfnamefont {F.}~\bibnamefont {Yang}},\ and\ \bibinfo {author} {\bibfnamefont {J.}~\bibnamefont {Guo}},\ }\href {https://doi.org/10.1063/5.0009481} {\bibfield  {journal} {\bibinfo  {journal} {Journal of Applied Physics}\ }\textbf {\bibinfo {volume} {128}},\ \bibinfo {pages} {035301} (\bibinfo {year} {2020})}\BibitemShut {NoStop}%
\bibitem [{\citenamefont {Biscaras}\ \emph {et~al.}(2010)\citenamefont {Biscaras}, \citenamefont {Bergeal}, \citenamefont {Kushwaha}, \citenamefont {Wolf}, \citenamefont {Rastogi}, \citenamefont {Budhani},\ and\ \citenamefont {Lesueur}}]{biscaras2010two}%
  \BibitemOpen
  \bibfield  {author} {\bibinfo {author} {\bibfnamefont {J.}~\bibnamefont {Biscaras}}, \bibinfo {author} {\bibfnamefont {N.}~\bibnamefont {Bergeal}}, \bibinfo {author} {\bibfnamefont {A.}~\bibnamefont {Kushwaha}}, \bibinfo {author} {\bibfnamefont {T.}~\bibnamefont {Wolf}}, \bibinfo {author} {\bibfnamefont {A.}~\bibnamefont {Rastogi}}, \bibinfo {author} {\bibfnamefont {R.~C.}\ \bibnamefont {Budhani}},\ and\ \bibinfo {author} {\bibfnamefont {J.}~\bibnamefont {Lesueur}},\ }\href {https://doi.org/10.1038/ncomms1084} {\bibfield  {journal} {\bibinfo  {journal} {Nature Communications}\ }\textbf {\bibinfo {volume} {1}},\ \bibinfo {pages} {89} (\bibinfo {year} {2010})}\BibitemShut {NoStop}%
\bibitem [{sm()}]{sm}%
  \BibitemOpen
  \bibinfo {note} {See Supplemental Material below.}\BibitemShut {Stop}%
\bibitem [{\citenamefont {Joshua}\ \emph {et~al.}(2012)\citenamefont {Joshua}, \citenamefont {Pecker}, \citenamefont {Ruhman}, \citenamefont {Altman},\ and\ \citenamefont {Ilani}}]{joshua2012universal}%
  \BibitemOpen
  \bibfield  {author} {\bibinfo {author} {\bibfnamefont {A.}~\bibnamefont {Joshua}}, \bibinfo {author} {\bibfnamefont {S.}~\bibnamefont {Pecker}}, \bibinfo {author} {\bibfnamefont {J.}~\bibnamefont {Ruhman}}, \bibinfo {author} {\bibfnamefont {E.}~\bibnamefont {Altman}},\ and\ \bibinfo {author} {\bibfnamefont {S.}~\bibnamefont {Ilani}},\ }\href {https://doi.org/10.1038/ncomms2116} {\bibfield  {journal} {\bibinfo  {journal} {Nature Communications}\ }\textbf {\bibinfo {volume} {3}},\ \bibinfo {pages} {1129} (\bibinfo {year} {2012})}\BibitemShut {NoStop}%
\bibitem [{\citenamefont {Maniv}\ \emph {et~al.}(2015)\citenamefont {Maniv}, \citenamefont {Shalom}, \citenamefont {Ron}, \citenamefont {Mograbi}, \citenamefont {Palevski}, \citenamefont {Goldstein},\ and\ \citenamefont {Dagan}}]{maniv2015strong}%
  \BibitemOpen
  \bibfield  {author} {\bibinfo {author} {\bibfnamefont {E.}~\bibnamefont {Maniv}}, \bibinfo {author} {\bibfnamefont {M.~B.}\ \bibnamefont {Shalom}}, \bibinfo {author} {\bibfnamefont {A.}~\bibnamefont {Ron}}, \bibinfo {author} {\bibfnamefont {M.}~\bibnamefont {Mograbi}}, \bibinfo {author} {\bibfnamefont {A.}~\bibnamefont {Palevski}}, \bibinfo {author} {\bibfnamefont {M.}~\bibnamefont {Goldstein}},\ and\ \bibinfo {author} {\bibfnamefont {Y.}~\bibnamefont {Dagan}},\ }\href {https://doi.org/10.1038/ncomms9239} {\bibfield  {journal} {\bibinfo  {journal} {Nature Communications}\ }\textbf {\bibinfo {volume} {6}},\ \bibinfo {pages} {8239} (\bibinfo {year} {2015})}\BibitemShut {NoStop}%
\bibitem [{\citenamefont {Khanna}\ \emph {et~al.}(2019)\citenamefont {Khanna}, \citenamefont {Rout}, \citenamefont {Mograbi}, \citenamefont {Tuvia}, \citenamefont {Leermakers}, \citenamefont {Zeitler}, \citenamefont {Dagan},\ and\ \citenamefont {Goldstein}}]{khanna2019symmetry}%
  \BibitemOpen
  \bibfield  {author} {\bibinfo {author} {\bibfnamefont {U.}~\bibnamefont {Khanna}}, \bibinfo {author} {\bibfnamefont {P.~K.}\ \bibnamefont {Rout}}, \bibinfo {author} {\bibfnamefont {M.}~\bibnamefont {Mograbi}}, \bibinfo {author} {\bibfnamefont {G.}~\bibnamefont {Tuvia}}, \bibinfo {author} {\bibfnamefont {I.}~\bibnamefont {Leermakers}}, \bibinfo {author} {\bibfnamefont {U.}~\bibnamefont {Zeitler}}, \bibinfo {author} {\bibfnamefont {Y.}~\bibnamefont {Dagan}},\ and\ \bibinfo {author} {\bibfnamefont {M.}~\bibnamefont {Goldstein}},\ }\href {https://doi.org/10.1103/PhysRevLett.123.036805} {\bibfield  {journal} {\bibinfo  {journal} {Phys. Rev. Lett.}\ }\textbf {\bibinfo {volume} {123}},\ \bibinfo {pages} {036805} (\bibinfo {year} {2019})}\BibitemShut {NoStop}%
\bibitem [{\citenamefont {Joshua}\ \emph {et~al.}(2013)\citenamefont {Joshua}, \citenamefont {Ruhman}, \citenamefont {Pecker}, \citenamefont {Altman},\ and\ \citenamefont {Ilani}}]{joshua2013gate}%
  \BibitemOpen
  \bibfield  {author} {\bibinfo {author} {\bibfnamefont {A.}~\bibnamefont {Joshua}}, \bibinfo {author} {\bibfnamefont {J.}~\bibnamefont {Ruhman}}, \bibinfo {author} {\bibfnamefont {S.}~\bibnamefont {Pecker}}, \bibinfo {author} {\bibfnamefont {E.}~\bibnamefont {Altman}},\ and\ \bibinfo {author} {\bibfnamefont {S.}~\bibnamefont {Ilani}},\ }\href {https://doi.org/10.1073/pnas.1221453110} {\bibfield  {journal} {\bibinfo  {journal} {Proceedings of the National Academy of Sciences}\ }\textbf {\bibinfo {volume} {110}},\ \bibinfo {pages} {9633} (\bibinfo {year} {2013})}\BibitemShut {NoStop}%
\end{thebibliography}%


\begin{thebibliography}{8}%
\makeatletter
\providecommand \@ifxundefined [1]{%
 \@ifx{#1\undefined}
}%
\providecommand \@ifnum [1]{%
 \ifnum #1\expandafter \@firstoftwo
 \else \expandafter \@secondoftwo
 \fi
}%
\providecommand \@ifx [1]{%
 \ifx #1\expandafter \@firstoftwo
 \else \expandafter \@secondoftwo
 \fi
}%
\providecommand \natexlab [1]{#1}%
\providecommand \enquote  [1]{``#1''}%
\providecommand \bibnamefont  [1]{#1}%
\providecommand \bibfnamefont [1]{#1}%
\providecommand \citenamefont [1]{#1}%
\providecommand \href@noop [0]{\@secondoftwo}%
\providecommand \href [0]{\begingroup \@sanitize@url \@href}%
\providecommand \@href[1]{\@@startlink{#1}\@@href}%
\providecommand \@@href[1]{\endgroup#1\@@endlink}%
\providecommand \@sanitize@url [0]{\catcode `\\12\catcode `\$12\catcode `\&12\catcode `\#12\catcode `\^12\catcode `\_12\catcode `\%12\relax}%
\providecommand \@@startlink[1]{}%
\providecommand \@@endlink[0]{}%
\providecommand \url  [0]{\begingroup\@sanitize@url \@url }%
\providecommand \@url [1]{\endgroup\@href {#1}{\urlprefix }}%
\providecommand \urlprefix  [0]{URL }%
\providecommand \Eprint [0]{\href }%
\providecommand \doibase [0]{https://doi.org/}%
\providecommand \selectlanguage [0]{\@gobble}%
\providecommand \bibinfo  [0]{\@secondoftwo}%
\providecommand \bibfield  [0]{\@secondoftwo}%
\providecommand \translation [1]{[#1]}%
\providecommand \BibitemOpen [0]{}%
\providecommand \bibitemStop [0]{}%
\providecommand \bibitemNoStop [0]{.\EOS\space}%
\providecommand \EOS [0]{\spacefactor3000\relax}%
\providecommand \BibitemShut  [1]{\csname bibitem#1\endcsname}%
\let\auto@bib@innerbib\@empty
\bibitem [{\citenamefont {Ideue}\ \emph {et~al.}(2017)\citenamefont {Ideue}, \citenamefont {Hamamoto}, \citenamefont {Koshikawa}, \citenamefont {Ezawa}, \citenamefont {Shimizu}, \citenamefont {Kaneko}, \citenamefont {Tokura}, \citenamefont {Nagaosa},\ and\ \citenamefont {Iwasa}}]{ideue2017bulk}%
  \BibitemOpen
  \bibfield  {author} {\bibinfo {author} {\bibfnamefont {T.}~\bibnamefont {Ideue}}, \bibinfo {author} {\bibfnamefont {K.}~\bibnamefont {Hamamoto}}, \bibinfo {author} {\bibfnamefont {S.}~\bibnamefont {Koshikawa}}, \bibinfo {author} {\bibfnamefont {M.}~\bibnamefont {Ezawa}}, \bibinfo {author} {\bibfnamefont {S.}~\bibnamefont {Shimizu}}, \bibinfo {author} {\bibfnamefont {Y.}~\bibnamefont {Kaneko}}, \bibinfo {author} {\bibfnamefont {Y.}~\bibnamefont {Tokura}}, \bibinfo {author} {\bibfnamefont {N.}~\bibnamefont {Nagaosa}},\ and\ \bibinfo {author} {\bibfnamefont {Y.}~\bibnamefont {Iwasa}},\ }\href {https://doi.org/10.1038/nphys4056} {\bibfield  {journal} {\bibinfo  {journal} {Nature Physics}\ }\textbf {\bibinfo {volume} {13}},\ \bibinfo {pages} {578} (\bibinfo {year} {2017})}\BibitemShut {NoStop}%
\bibitem [{\citenamefont {Battilomo}\ \emph {et~al.}(2021)\citenamefont {Battilomo}, \citenamefont {Scopigno},\ and\ \citenamefont {Ortix}}]{battilomo2021anomalous}%
  \BibitemOpen
  \bibfield  {author} {\bibinfo {author} {\bibfnamefont {R.}~\bibnamefont {Battilomo}}, \bibinfo {author} {\bibfnamefont {N.}~\bibnamefont {Scopigno}},\ and\ \bibinfo {author} {\bibfnamefont {C.}~\bibnamefont {Ortix}},\ }\href {https://doi.org/10.1103/PhysRevResearch.3.L012006} {\bibfield  {journal} {\bibinfo  {journal} {Phys. Rev. Res.}\ }\textbf {\bibinfo {volume} {3}},\ \bibinfo {pages} {L012006} (\bibinfo {year} {2021})}\BibitemShut {NoStop}%
\bibitem [{\citenamefont {He}\ \emph {et~al.}(2018)\citenamefont {He}, \citenamefont {Walker}, \citenamefont {Zhang}, \citenamefont {Bruno}, \citenamefont {Bahramy}, \citenamefont {Lee}, \citenamefont {Ramaswamy}, \citenamefont {Cai}, \citenamefont {Heinonen}, \citenamefont {Vignale}, \citenamefont {Baumberger},\ and\ \citenamefont {Yang}}]{he2018observation}%
  \BibitemOpen
  \bibfield  {author} {\bibinfo {author} {\bibfnamefont {P.}~\bibnamefont {He}}, \bibinfo {author} {\bibfnamefont {S.~M.}\ \bibnamefont {Walker}}, \bibinfo {author} {\bibfnamefont {S.~S.-L.}\ \bibnamefont {Zhang}}, \bibinfo {author} {\bibfnamefont {F.~Y.}\ \bibnamefont {Bruno}}, \bibinfo {author} {\bibfnamefont {M.~S.}\ \bibnamefont {Bahramy}}, \bibinfo {author} {\bibfnamefont {J.~M.}\ \bibnamefont {Lee}}, \bibinfo {author} {\bibfnamefont {R.}~\bibnamefont {Ramaswamy}}, \bibinfo {author} {\bibfnamefont {K.}~\bibnamefont {Cai}}, \bibinfo {author} {\bibfnamefont {O.}~\bibnamefont {Heinonen}}, \bibinfo {author} {\bibfnamefont {G.}~\bibnamefont {Vignale}}, \bibinfo {author} {\bibfnamefont {F.}~\bibnamefont {Baumberger}},\ and\ \bibinfo {author} {\bibfnamefont {H.}~\bibnamefont {Yang}},\ }\href {https://doi.org/10.1103/PhysRevLett.120.266802} {\bibfield  {journal} {\bibinfo  {journal} {Phys. Rev. Lett.}\ }\textbf {\bibinfo {volume} {120}},\ \bibinfo {pages} {266802} (\bibinfo {year} {2018})}\BibitemShut {NoStop}%
\bibitem [{\citenamefont {Rout}\ \emph {et~al.}(2017)\citenamefont {Rout}, \citenamefont {Maniv},\ and\ \citenamefont {Dagan}}]{rout2017link}%
  \BibitemOpen
  \bibfield  {author} {\bibinfo {author} {\bibfnamefont {P.~K.}\ \bibnamefont {Rout}}, \bibinfo {author} {\bibfnamefont {E.}~\bibnamefont {Maniv}},\ and\ \bibinfo {author} {\bibfnamefont {Y.}~\bibnamefont {Dagan}},\ }\href {https://doi.org/10.1103/PhysRevLett.119.237002} {\bibfield  {journal} {\bibinfo  {journal} {Phys. Rev. Lett.}\ }\textbf {\bibinfo {volume} {119}},\ \bibinfo {pages} {237002} (\bibinfo {year} {2017})}\BibitemShut {NoStop}%
\bibitem [{\citenamefont {Lesne}\ \emph {et~al.}(2023)\citenamefont {Lesne}, \citenamefont {Sa{\v{g}}lam}, \citenamefont {Battilomo}, \citenamefont {Mercaldo}, \citenamefont {van Thiel}, \citenamefont {Filippozzi}, \citenamefont {Noce}, \citenamefont {Cuoco}, \citenamefont {Steele}, \citenamefont {Ortix},\ and\ \citenamefont {Caviglia}}]{lesne2023designing}%
  \BibitemOpen
  \bibfield  {author} {\bibinfo {author} {\bibfnamefont {E.}~\bibnamefont {Lesne}}, \bibinfo {author} {\bibfnamefont {Y.~G.}\ \bibnamefont {Sa{\v{g}}lam}}, \bibinfo {author} {\bibfnamefont {R.}~\bibnamefont {Battilomo}}, \bibinfo {author} {\bibfnamefont {M.~T.}\ \bibnamefont {Mercaldo}}, \bibinfo {author} {\bibfnamefont {T.~C.}\ \bibnamefont {van Thiel}}, \bibinfo {author} {\bibfnamefont {U.}~\bibnamefont {Filippozzi}}, \bibinfo {author} {\bibfnamefont {C.}~\bibnamefont {Noce}}, \bibinfo {author} {\bibfnamefont {M.}~\bibnamefont {Cuoco}}, \bibinfo {author} {\bibfnamefont {G.~A.}\ \bibnamefont {Steele}}, \bibinfo {author} {\bibfnamefont {C.}~\bibnamefont {Ortix}},\ and\ \bibinfo {author} {\bibfnamefont {A.~D.}\ \bibnamefont {Caviglia}},\ }\href {https://doi.org/10.1038/s41563-023-01498-0} {\bibfield  {journal} {\bibinfo  {journal} {Nature Materials}\ }\textbf {\bibinfo {volume} {22}},\ \bibinfo {pages} {576} (\bibinfo {year} {2023})}\BibitemShut {NoStop}%
\bibitem [{\citenamefont {Ben~Shalom}\ \emph {et~al.}(2010)\citenamefont {Ben~Shalom}, \citenamefont {Sachs}, \citenamefont {Rakhmilevitch}, \citenamefont {Palevski},\ and\ \citenamefont {Dagan}}]{shalom2010tuning}%
  \BibitemOpen
  \bibfield  {author} {\bibinfo {author} {\bibfnamefont {M.}~\bibnamefont {Ben~Shalom}}, \bibinfo {author} {\bibfnamefont {M.}~\bibnamefont {Sachs}}, \bibinfo {author} {\bibfnamefont {D.}~\bibnamefont {Rakhmilevitch}}, \bibinfo {author} {\bibfnamefont {A.}~\bibnamefont {Palevski}},\ and\ \bibinfo {author} {\bibfnamefont {Y.}~\bibnamefont {Dagan}},\ }\href {https://doi.org/10.1103/PhysRevLett.104.126802} {\bibfield  {journal} {\bibinfo  {journal} {Phys. Rev. Lett.}\ }\textbf {\bibinfo {volume} {104}},\ \bibinfo {pages} {126802} (\bibinfo {year} {2010})}\BibitemShut {NoStop}%
\bibitem [{\citenamefont {Joshua}\ \emph {et~al.}(2013)\citenamefont {Joshua}, \citenamefont {Ruhman}, \citenamefont {Pecker}, \citenamefont {Altman},\ and\ \citenamefont {Ilani}}]{joshua2013gate}%
  \BibitemOpen
  \bibfield  {author} {\bibinfo {author} {\bibfnamefont {A.}~\bibnamefont {Joshua}}, \bibinfo {author} {\bibfnamefont {J.}~\bibnamefont {Ruhman}}, \bibinfo {author} {\bibfnamefont {S.}~\bibnamefont {Pecker}}, \bibinfo {author} {\bibfnamefont {E.}~\bibnamefont {Altman}},\ and\ \bibinfo {author} {\bibfnamefont {S.}~\bibnamefont {Ilani}},\ }\href {https://doi.org/10.1073/pnas.1221453110} {\bibfield  {journal} {\bibinfo  {journal} {Proceedings of the National Academy of Sciences}\ }\textbf {\bibinfo {volume} {110}},\ \bibinfo {pages} {9633} (\bibinfo {year} {2013})}\BibitemShut {NoStop}%
\bibitem [{\citenamefont {Maniv}\ \emph {et~al.}(2015)\citenamefont {Maniv}, \citenamefont {Shalom}, \citenamefont {Ron}, \citenamefont {Mograbi}, \citenamefont {Palevski}, \citenamefont {Goldstein},\ and\ \citenamefont {Dagan}}]{maniv2015strong}%
  \BibitemOpen
  \bibfield  {author} {\bibinfo {author} {\bibfnamefont {E.}~\bibnamefont {Maniv}}, \bibinfo {author} {\bibfnamefont {M.~B.}\ \bibnamefont {Shalom}}, \bibinfo {author} {\bibfnamefont {A.}~\bibnamefont {Ron}}, \bibinfo {author} {\bibfnamefont {M.}~\bibnamefont {Mograbi}}, \bibinfo {author} {\bibfnamefont {A.}~\bibnamefont {Palevski}}, \bibinfo {author} {\bibfnamefont {M.}~\bibnamefont {Goldstein}},\ and\ \bibinfo {author} {\bibfnamefont {Y.}~\bibnamefont {Dagan}},\ }\href {https://doi.org/10.1038/ncomms9239} {\bibfield  {journal} {\bibinfo  {journal} {Nature Communications}\ }\textbf {\bibinfo {volume} {6}},\ \bibinfo {pages} {8239} (\bibinfo {year} {2015})}\BibitemShut {NoStop}%
\end{thebibliography}%
\end{document}